\documentclass[AMA,Times1COL]{WileyNJDv5} %STIX1COL,STIX2COL,STIXSMALL

\articletype{Article Type}%

\received{Date Month Year}
\revised{Date Month Year}
\accepted{Date Month Year}
\journal{Journal}
\volume{00}
\copyyear{2023}
\startpage{1}

\begin{document}

\title{Opt4GPTQ: Co-Optimizing Memory and Computation for 4-bit GPTQ Quantized LLM Inference on Heterogeneous Platforms}

\author[1]{Yaozheng Zhang$\dagger$}

\author[1]{Wei Wang$\dagger$}

\author[1]{Jie Kong}
\author[1]{Jiehan Zhou*}
\author[2]{Xianwei Zhang}
\author[1]{Huanqing Cui}
\author[1]{Han Bao}
\author[3]{Yuhai Liu}

\authormark{TAYLOR \textsc{et al.}}
\titlemark{PLEASE INSERT YOUR ARTICLE TITLE HERE}

\address[1]{\orgdiv{School of Computer Science and Engineering}, \orgname{Shandong University of Science and Technology}, \orgaddress{\state{Qingdao}, \country{China}}}
\address[2]{\orgdiv{School of Computer Science and Engineering}, \orgname{Sun Yet-sen University}, \orgaddress{\state{Guangzhou}, \country{China}}}
\address[3]{\orgname{Dawning International Information Industry Co., Ltd.}, \orgaddress{\state{Qingdao}, \country{China}}}

\corres{Corresponding author: Jiehan Zhou, \email{jiehan.zhou@sdust.edu.cn}}

\presentaddress{$\dagger$These authors contributed equally to this work.}

%\fundingInfo{Text}
%\JELinfo{ejlje}

\abstract[Abstract]{The increasing adoption of large language models (LLMs) on heterogeneous computing platforms poses significant challenges to achieving high inference efficiency. To address these efficiency bottlenecks across diverse platforms, this paper proposes Opt4GPTQ, a practical optimization method designed for 4-bit GPTQ quantized LLMs inference on heterogeneous AI accelerators. Built upon the vLLM serving system, Opt4GPTQ integrates three platform-level optimization strategies: Shared Memory Buffering Optimization (SMB-Opt), which caches frequently accessed data in shared memory and employs single-threaded writes; Vectorized Memory Loading Optimization (VML-Opt), which utilizes vectorized memory operations for efficient data loading; and Inline Assembly Optimization (ILA-Opt), which directly leverages hardware-native vector half-precision addition and fused multiply-accumulate instructions. Experimental results show that Opt4GPTQ effectively improves performance across various models while maintaining original model accuracy, achieving throughput gains of up to 84.42\%. This work highlights the critical role of platform-level engineering in enabling efficient LLMs inference on emerging architectures and provides valuable methodologies for future heterogeneous platform adaptation.}


\jnlcitation{\cname{%
\author{Taylor M.},
\author{Lauritzen P},
\author{Erath C}, and
\author{Mittal R}}.
\ctitle{On simplifying ‘incremental remap’-based transport schemes.} \cjournal{\it J Comput Phys.} \cvol{2021;00(00):1--18}.}

\maketitle

\renewcommand\thefootnote{}

\renewcommand\thefootnote{\fnsymbol{footnote}}
\setcounter{footnote}{1}

\section{Introduction}\label{sec1}

Large Language Models (LLMs), such as Llama \cite{b1}, ChatGPT \cite{b2}, DeepSeek-V2 \cite{b3}, and Gemini \cite{b4}, have demonstrated broad application potential in tasks like conversational agents and text generation. However, their substantial parameter scales, coupled with high computational loads, memory access overhead, and memory consumption during inference, present significant challenges for achieving efficient execution. These issues are particularly pronounced in resource-constrained edge environments and remain critical in cloud scenarios where performance and energy efficiency are paramount. Consequently, enhancing inference efficiency while maintaining model accuracy has become a key focus of current research \cite{b5,b6}.

To address these challenges, the vLLM serving system was proposed \cite{b7}. vLLM optimizes key-value (KV) cache management by flexibly sharing KV caches across single and multiple requests, thereby reducing overall memory usage. In addition, vLLM supports various low-bit quantization techniques, leveraging low-precision weights and activations combined with optimized compute cores to accelerate inference \cite{b8,b9}. Among weight-only quantization methods, Generative Pre-trained Transformer Quantization (GPTQ) has attracted attention for its ability to achieve 4-bit quantization while maintaining acceptable inference accuracy \cite{b10}. GPTQ combines one-shot weight quantization with approximate second-order information, enabling large-scale LLMs to be compressed to 4-bit precision in a relatively short time with minimal accuracy degradation, making it one of the mainstream low-bit quantization approaches.

Although vLLM and GPTQ have achieved mature performance optimizations on conventional accelerator platforms such as GPUs, the emergence of heterogeneous computing platforms introduces new hardware architectures that pose additional adaptation challenges. By integrating CPUs with GPUs, NPUs, or other domain-specific accelerators, heterogeneous platforms offer new opportunities to enhance LLM inference efficiency \cite{b11,b12}. However, significant architectural differences among platforms hinder the direct deployment of GPU-optimized inference frameworks on these accelerators, making hardware-specific adaptation and performance tuning critical issues to address.

The Deep Computing Unit (DCU) of HYGON is a heterogeneous accelerator designed for AI workloads \cite{b13}. It adopts a general-purpose GPGPU architecture and maintains full compatibility with the ROCm ecosystem, theoretically ensuring CUDA compatibility. However, due to insufficient exploitation of the underlying hardware architecture by current inference engines, the inference performance on DCU has not yet reached its expected potential. In particular, in quantized inference scenarios, issues such as throughput degradation and increased latency significantly limit the performance scalability of LLM inference on DCU.

To tackle these challenges, this paper presents a systematic engineering optimization and adaptation study of the 4-bit GPTQ quantized inference in the vLLM serving system on the DCU heterogeneous accelerator platform, proposing the Opt4GPTQ approach. This approach integrates three optimization strategies: Shared Memory Buffering Optimization (SMB-Opt), which caches data in shared memory and employs single-threaded writes to reduce global memory access overhead; Vectorized Memory Loading Optimization (VML-Opt), which utilizes vectorized memory operations for efficient data loading; and Inline Assembly Optimization (ILA-Opt), which directly utilizes the hardware's native vectorized half-precision addition and fused multiply-add instructions for efficient execution. The proposed optimization methods and engineering practices provide valuable references for LLM inference optimization on other emerging heterogeneous accelerator platforms as well.

The remainder of this paper is organized as follows: Section 2 introduces related work. Section 3 introduces Opt4GPTQ. Section 4 presents the experimental setup and results evaluation. Finally, Section 5 concludes the paper and discusses future research directions.

\section{Related Work}\label{sec2}

\subsection{Atomic Operation Synchronization}\label{subsec1}
Parallel reduction within quantized GEMM kernels heavily relies on atomic memory operations (AMOs), which frequently introduce serialization delays. Detailed characterizations of heterogeneous systems reveal that atomic updates on floating-point data often rely on inefficient compare-and-swap (CAS) constructs, leading to significant cache coherence traffic \cite{b25}. Research comparing "near" AMOs (executed in private caches) and "far" AMOs (executed in memory) suggests that performing synchronization closer to the compute units can drastically improve throughput. Furthermore, modeling the utilization of shared-memory atomic units has identified that even within shared memory, contention can vary significantly depending on the access pattern \cite{b26}. To mitigate these bottlenecks, our SMB-Opt adopts a hierarchical buffering strategy. By caching intermediate sums in shared memory and employing single-threaded global writes, we effectively minimize the "far" AMOs overheads and global memory contention highlighted in the literature.

\subsection{Memory Access Patterns and Vectorization}\label{subsec2}

Efficient memory access is a primary bottleneck for high-throughput LLM inference on heterogeneous platforms. Recent studies on automatic data layout transformation indicate that optimizing memory structures—such as transforming data into a structure-of-arrays—can significantly enhance cache utilization and facilitate auto-vectorization \cite{b27}. However, conventional vector processors often encounter substantial efficiency gaps when handling strided or segmented memory access patterns \cite{b28}. To bridge this gap, architectural optimizations have proposed shifting-based networks to coalesce irregular accesses into contiguous streams. Building on these principles, our VML-Opt implements software-level vectorized loading by explicitly managing data packing. This ensures coalesced access to global memory without requiring the hardware-level modifications suggested in prior architectural research.

\subsection{Kernel-Level Instruction Optimization}\label{subsec3}

To maximize compute density, recent efforts have focused on fine-grained control over GPU work partitioning and parallelism. Optimizing thread block and warp-level coordination has been proven essential to reduce unnecessary shared memory I/O and increase occupancy \cite{b29}. While domain-specific languages like Triton attempt to automate kernel generation, benchmarking efforts show that automated compilers often fail to produce performance-optimized code for complex operators, lacking awareness of specific hardware Instruction Set Architecture (ISA) requirements \cite{b30}. This performance gap is particularly evident on non-CUDA platforms where compilers may generate redundant instruction streams. Our ILA-Opt addresses this by bypassing high-level abstractions and directly utilizing inline assembly to invoke hardware-native vector instructions. This manual instruction-level alignment ensures optimal register allocation and pipeline utilization, surpassing the efficiency of the automated approaches discussed in recent benchmarks.

\section{Method}\label{sec3}

\begin{figure}[!ht]
\centering
\includegraphics[width=0.9\columnwidth]{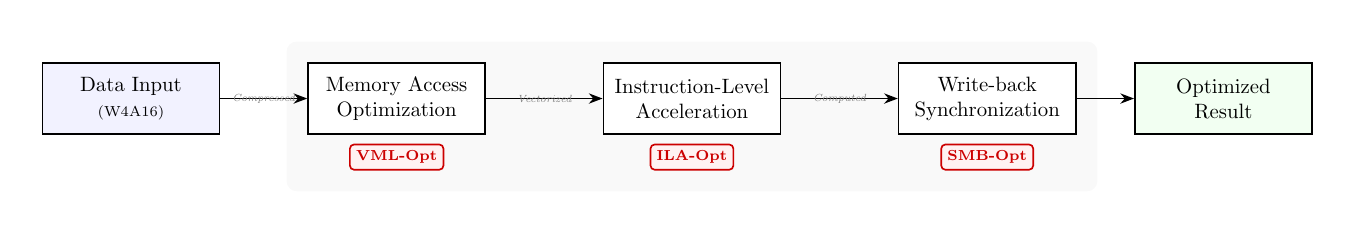}
\caption{Integrated Optimization Pipeline}
\label{fig3}
\end{figure}

As illustrated in Figure ~\ref{fig3}, Opt4GPTQ enhances the 4-bit GPTQ inference process by integrating three specific optimizations into the kernel execution pipeline: it leverages VML-Opt to aggregate fragmented memory accesses into vectorized loads via data re-interpretation; employs ILA-Opt to invoke hardware-native instructions through manual inline assembly for peak arithmetic throughput; and utilizes SMB-Opt to buffer intermediate results in shared memory, effectively resolving global memory contention by replacing massive atomic operations with a coordinated hierarchical reduction.

\subsection{SMB-Opt}\label{subsec1}

\begin{figure}[!ht]
\centering
\includegraphics[width=0.9\columnwidth]{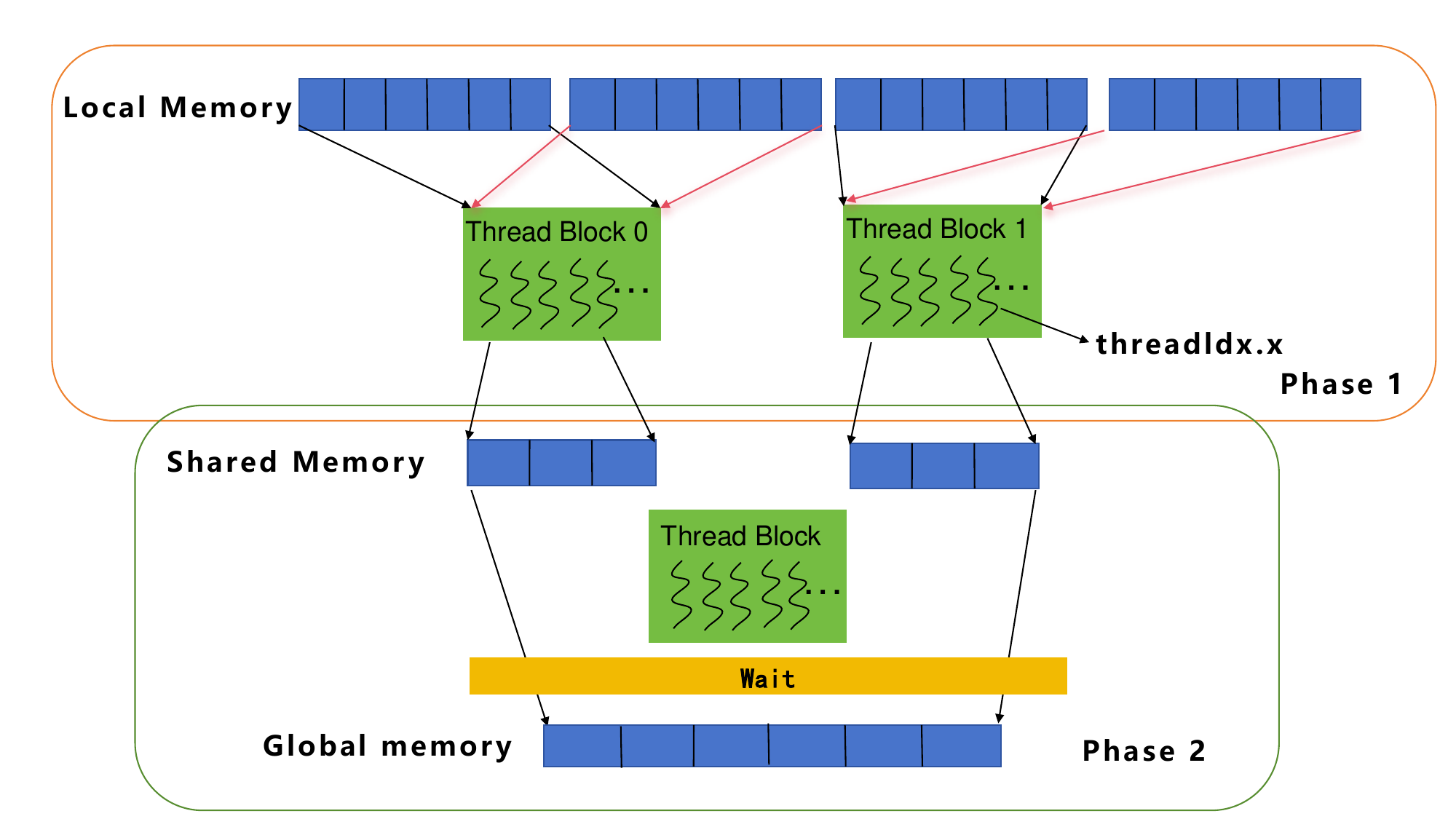}
\caption{Atomic operation optimization (SMB-Opt)}
\label{fig1}
\end{figure}

The SMB-Opt strategy is designed to resolve the performance bottleneck caused by massive atomic collisions and high-latency global memory access during the reduction phase. The overall logic is to introduce a hierarchical reduction mechanism using low-latency shared memory as an intermediate buffer. As illustrated in Figure~\ref{fig1}, the process is decoupled into two phases: in Phase 1, threads within a block accumulate partial results into a shared memory segment; in Phase 2, a single designated thread commits these aggregated results to the global memory via \texttt{atomicAdd}. This approach effectively transforms thousands of global memory contentions into block-level local operations, significantly enhancing write-back throughput.

The technical implementation is detailed in Algorithm~\ref{alg:smb-opt}. For each thread block, a shared memory array \textit{block\_result} is allocated to cache intermediate sums. Each thread calculates its partial result and performs intra-block accumulation, as expressed in Equations~\ref{eq6} and~\ref{eq7}:

\begin{equation}
\text{block\_result}[0] = \sum_{m=0}^{M-1} \text{(result01)}
\label{eq6}
\end{equation}

\begin{equation}
\text{block\_result}[1] = \sum_{m=0}^{M-1} \text{(result23)}
\label{eq7}
\end{equation}

To ensure data consistency before the final commitment, a synchronization barrier (\texttt{\_\_syncthreads()}) is enforced. Subsequently, the leading thread of each block retrieves the starting global memory address \textit{out} and performs the synchronized update, defined by Equations~(\ref{eq8}) through~(\ref{eq10}):

\begin{equation}
\text{out} = c\_.\texttt{item\_ptr}(\text{offset\_m}, n)
\label{eq8}
\end{equation}

\begin{equation}
\texttt{atomicAdd}(\text{out}, \text{block\_result}[0])
\label{eq9}
\end{equation}

\begin{equation}
\texttt{atomicAdd}(\text{out} + 1, \text{block\_result}[1])
\label{eq10}
\end{equation}

\begin{algorithm}[h]
\caption{SMB-Opt Algorithm}\label{alg:smb-opt}
\begin{algorithmic}[1]
\Require Matrix \textit{block\_c} of size $m \times n$
\Ensure Accumulated results in \textit{block\_result[0, 1]}
\State Define shared memory array \textit{block\_result[2]}
\If{\textit{threadIdx.x} == 0}
    \State Initialize \textit{block\_result} to \texttt{make\_half2(0, 0)}
\EndIf
\State \texttt{\_\_syncthreads()}
\For{each row \textit{m} in \textit{block\_c}}
    \State Accumulate values into \textit{block\_result} as \texttt{half2}
\EndFor
\State \texttt{\_\_syncthreads()}
\If{\textit{threadIdx.x} == 0}
    \State \texttt{atomicAdd}(\textit{out}, \textit{block\_result}) \Comment{Write to global memory}
\EndIf
\end{algorithmic}
\end{algorithm}

\subsection{VML-Opt}\label{subsec2}

VML-Opt enhances memory throughput by maximizing the utilization of the hardware's memory bus width. In contrast to standard sequential loading of 16-bit \texttt{half} elements, VML-Opt employs vectorized access to read two \texttt{half} elements simultaneously. This optimization reduces the total number of load instructions issued and improves the coalescing efficiency of global memory transactions.

As shown in Algorithm~\ref{alg:vml-opt}, during the loading of input matrix $\mathit{a}$, we utilize a \texttt{reinterpret\_cast} to treat the memory address as a \texttt{half2} structure, allowing the hardware to fetch two elements in a single clock cycle, as shown in Equation~(\ref{eq11}):

\begin{equation}
a_{0}\_h2 = *reinterpret\_cast<half2*>(\&a\_ptr[m])
\label{eq11}
\end{equation}

Once the 32-bit data is fetched into the register $a_{0}\_h2$, it is decomposed into two 16-bit \texttt{half} elements to be stored in the shared memory buffer, as described in Equations~(\ref{eq12}) and~(\ref{eq13}):

\begin{equation}
\text{block\_a\_ptr}(t) = \texttt{\_\_low2half}\left(a_{0}\_h2\right)
\label{eq12}
\end{equation}

\begin{equation}
\text{block\_a\_ptr}(t+1) = \texttt{\_\_high2half}\left(a_{0}\_h2\right)
\label{eq13}
\end{equation}

\begin{algorithm}[h]
\caption{VML-Opt Algorithm}\label{alg:vml-opt}
\begin{algorithmic}[1]
\Require Matrix \textit{a\_}, offsets, thread index \textit{t}
\Ensure Loaded data in shared memory \textit{block\_a}
\If{within bounds}
    \For{\textit{m} = 0 \textbf{to} \textit{m\_count} $- 1$}
        \State $a_{0}\_h2 \gets$ Load 32-bit \texttt{half2} from \textit{a\_ptr}
        \State \textit{block\_a\_ptr[t]} $\gets \texttt{\_\_low2half}(a_{0}\_h2)$
        \State \textit{block\_a\_ptr[t+1]} $\gets \texttt{\_\_high2half}(a_{0}\_h2)$
    \EndFor
\EndIf
\end{algorithmic}
\end{algorithm}

\subsection{ILA-Opt}\label{subsec3}

\begin{algorithm}[h]
\caption{ILA-Opt Algorithm}\label{alg:ila-opt}
\begin{algorithmic}[1]
\Function{hip\_hfma2}{$a, b, c$}
    \State \textbf{asm}: \texttt{v\_mad\_f16 \%0, \%1, \%2, \%3} \Comment{SIMO Fused Multiply-Add}
    \State \Return result
\EndFunction
\Function{hip\_hadd2}{$a, b$}
    \State \textbf{asm}: \texttt{v\_add\_f16 \%0, \%1, \%2} \Comment{SIMO Addition}
    \State \Return result
\EndFunction
\end{algorithmic}
\end{algorithm}

ILA-Opt focuses on instruction-level efficiency by directly mapping high-level operations to the hardware-native GCN/VOP3 instruction set of the HYGON DCU. By replacing built-in functions with custom inline assembly, ILA-Opt achieves single-instruction multiple-operation execution. This strategy minimizes compiler-induced overhead and ensures vector general-purpose register residency.

The optimization replaces high-level arithmetic calls with direct hardware primitives as shown in Algorithm~\ref{alg:ila-opt}. The customized \texttt{hip\_hfma2} function invokes the \texttt{v\_mad\_f16} instruction, enabling simultaneous fused multiply-add operations on two \texttt{half} elements within a single instruction, as defined in Equation~(\ref{eq14}):

\begin{equation}
r = a \times b + c
\label{eq14}
\end{equation}

Similarly, the \texttt{hip\_hadd2} function implements the \texttt{v\_add\_f16} instruction to perform vectorized addition directly at the hardware level, as shown in Equation~(\ref{eq15}):

\begin{equation}
r = a + b
\label{eq15}
\end{equation}

\section{Experiments}
\subsection{Experimental environment}
The experimental platform based on the HYGON DCU Z100 heterogeneous accelerator, and the ShareGPT\_V3\_unfiltered\_cleaned\_split dataset \cite{b17} is employed for throughput evaluation. The ARC \cite{b18} dataset is employed for accuracy evaluation. The ARC question set is partitioned into a Challenge Set (ARC\_C) and an Easy Set (ARC\_E). To enhance the robustness of the method, six models are used - Meta-Llama-3-8B-GPTQ \cite{b19}, Llama-2-7B-GPTQ \cite{b20}, CodeLlama-7B-GPTQ \cite{b21}, LLaMa-13B-GPTQ, Qwen1.5-4B-Chat-GPTQ-Int4 \cite{b22} and Qwen1.5-1.8B-Chat-GPTQ-Int4, while the generation throughput and inference accuracy of vLLM are used as the main performance indicators \cite{b23}. 

To ensure the reliability of the experimental results and minimize the impact of randomness and errors, the study repeated the experiment 15 times to enhance the scientific rigor of the findings. The final results were presented as the average of the 15 experimental trials.

\subsection{Baseline}

This study establishes a performance evaluation baseline using the vLLM serving system. It is noteworthy that the native GPTQ implementation in vLLM already incorporates industry-leading inference kernels, such as those derived from the ExLlamaV2 \cite{b31} project, to handle 4-bit quantization. These kernels represent a highly competitive state-of-the-art (SOTA) reference for efficiency analysis in large-scale serving scenarios. Our Opt4GPTQ framework is built upon this SOTA-level baseline to further address specific memory and computation bottlenecks identified on the HYGON DCU Z100 heterogeneous accelerator. In this configuration, input weights and activations are processed in FP16 format, and each evaluation run uses batch size of 32 prompts to ensure a consistent and rigorous performance assessment.

\subsection{Throughput Evaluation}

\begin{figure*}[htbp]
\centering
\includegraphics[width=0.48\textwidth]{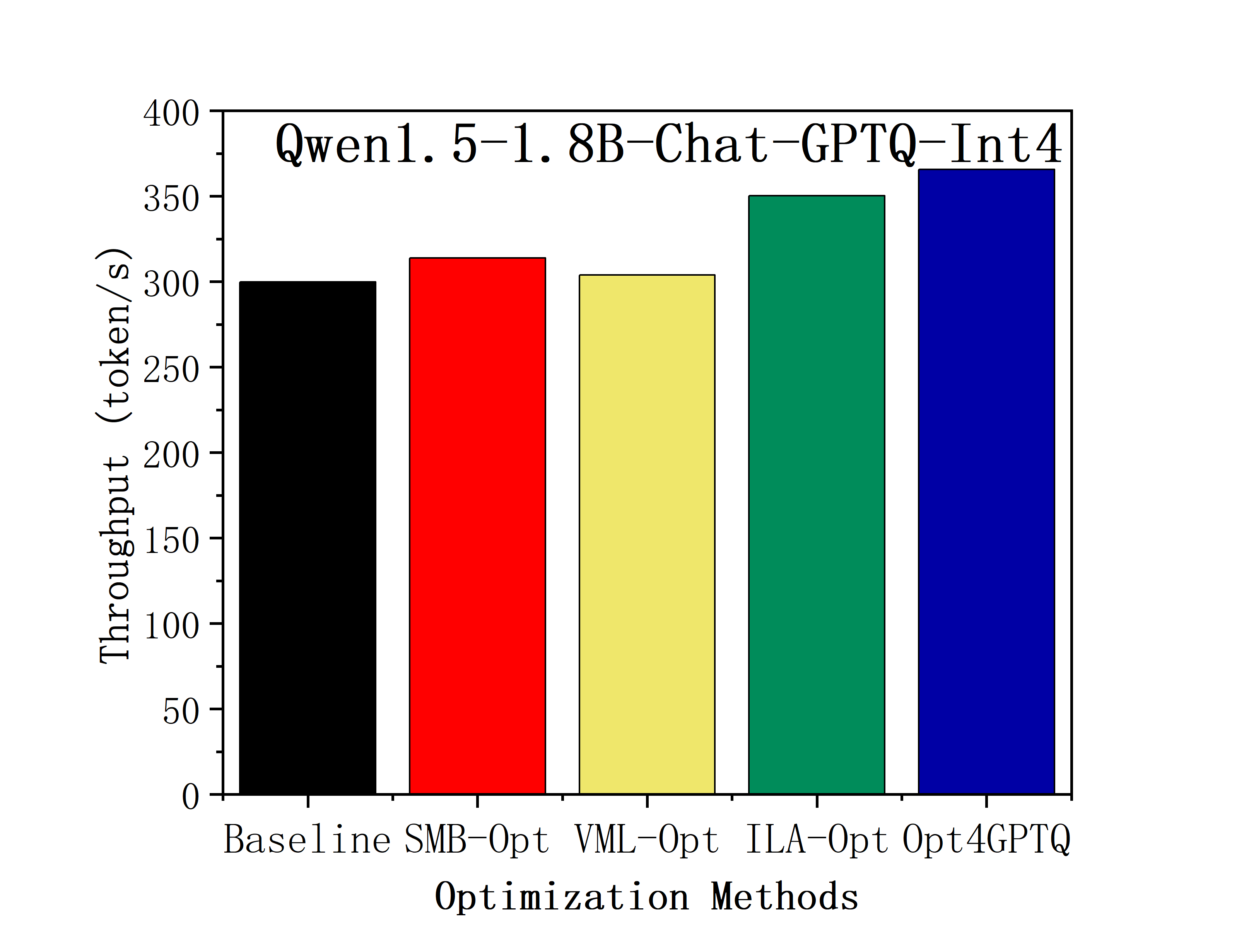} \hfill
\includegraphics[width=0.48\textwidth]{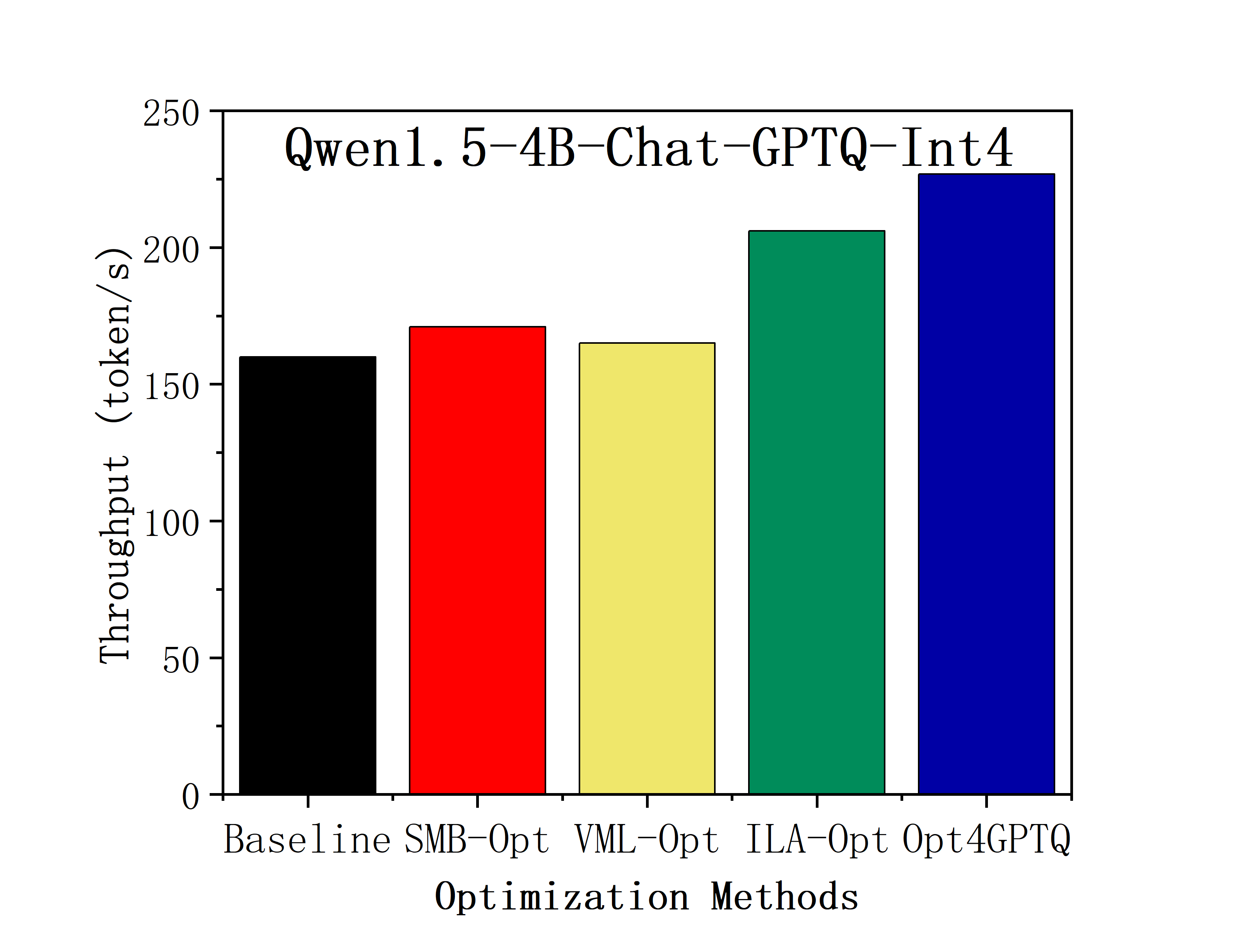} \\ \vspace{0.5cm}
\includegraphics[width=0.48\textwidth]{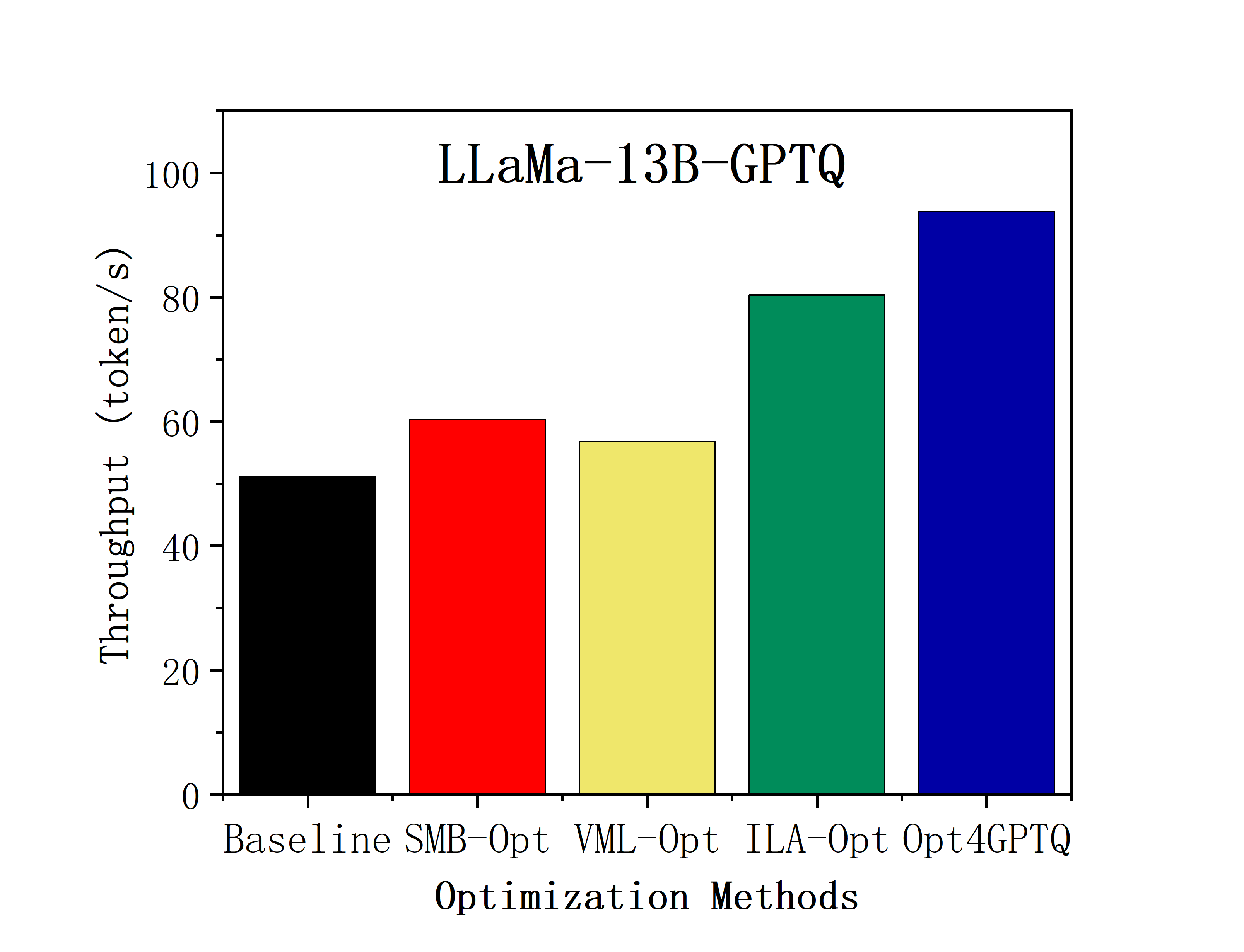} \hfill
\includegraphics[width=0.48\textwidth]{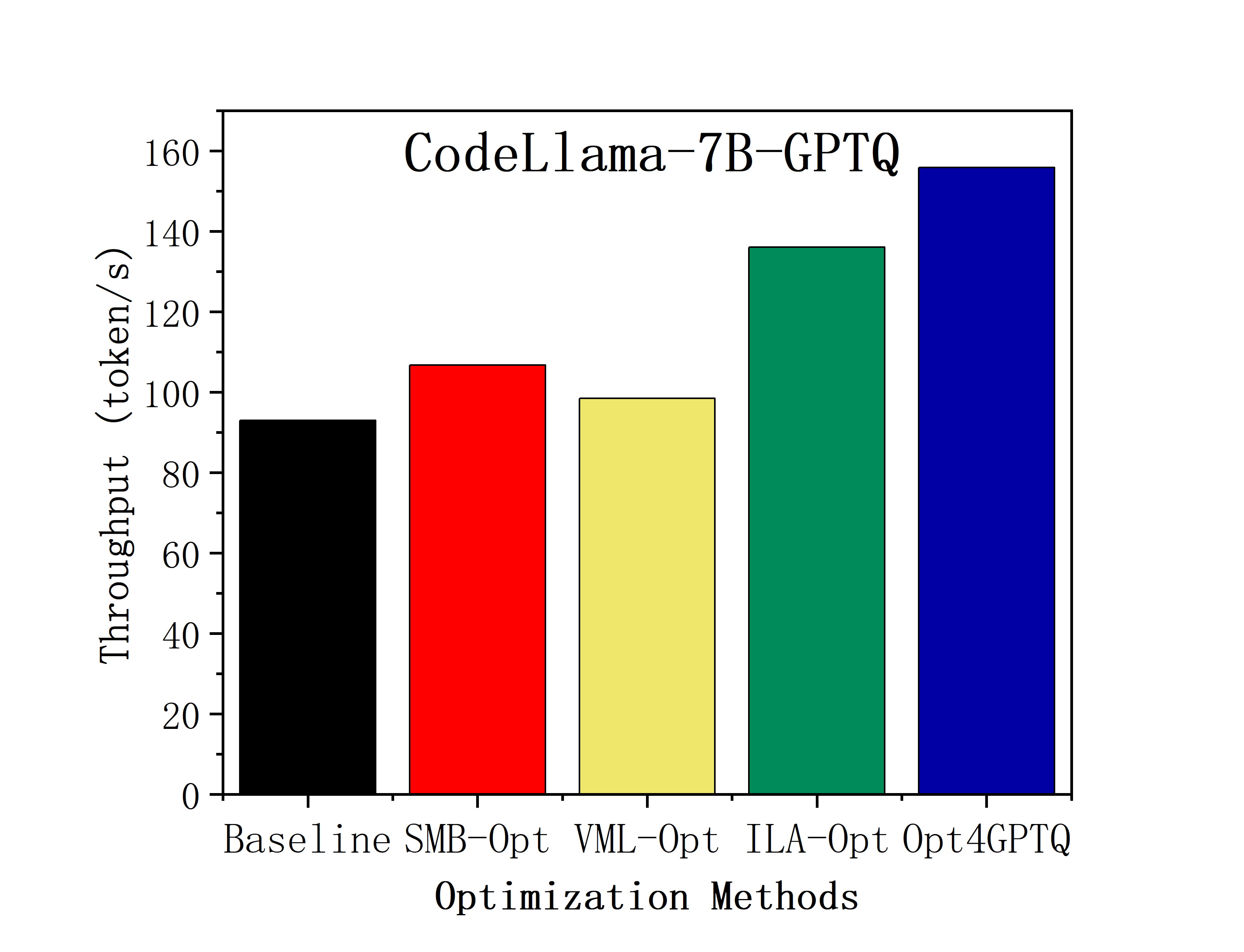} \\ \vspace{0.5cm}
\includegraphics[width=0.48\textwidth]{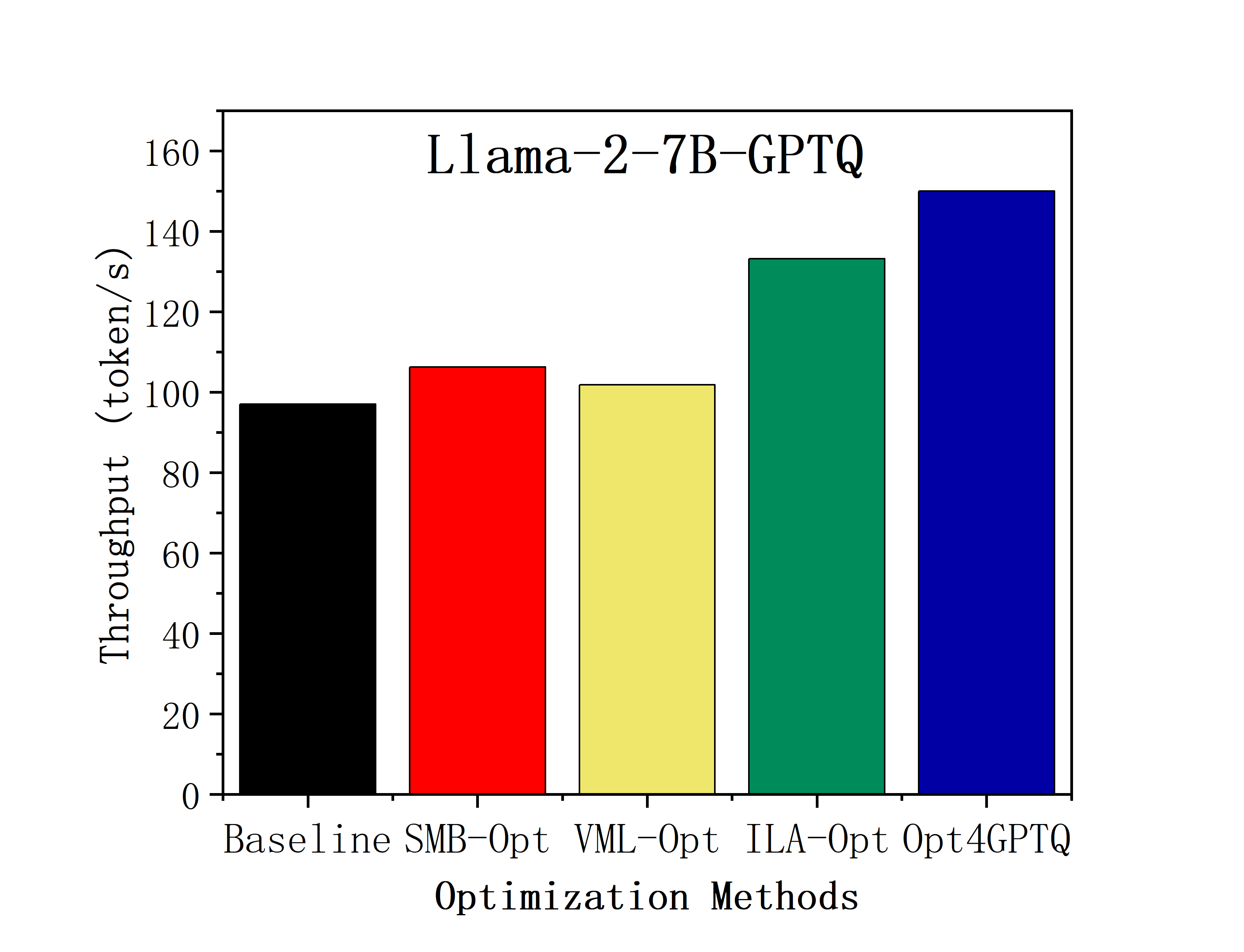} \hfill
\includegraphics[width=0.48\textwidth]{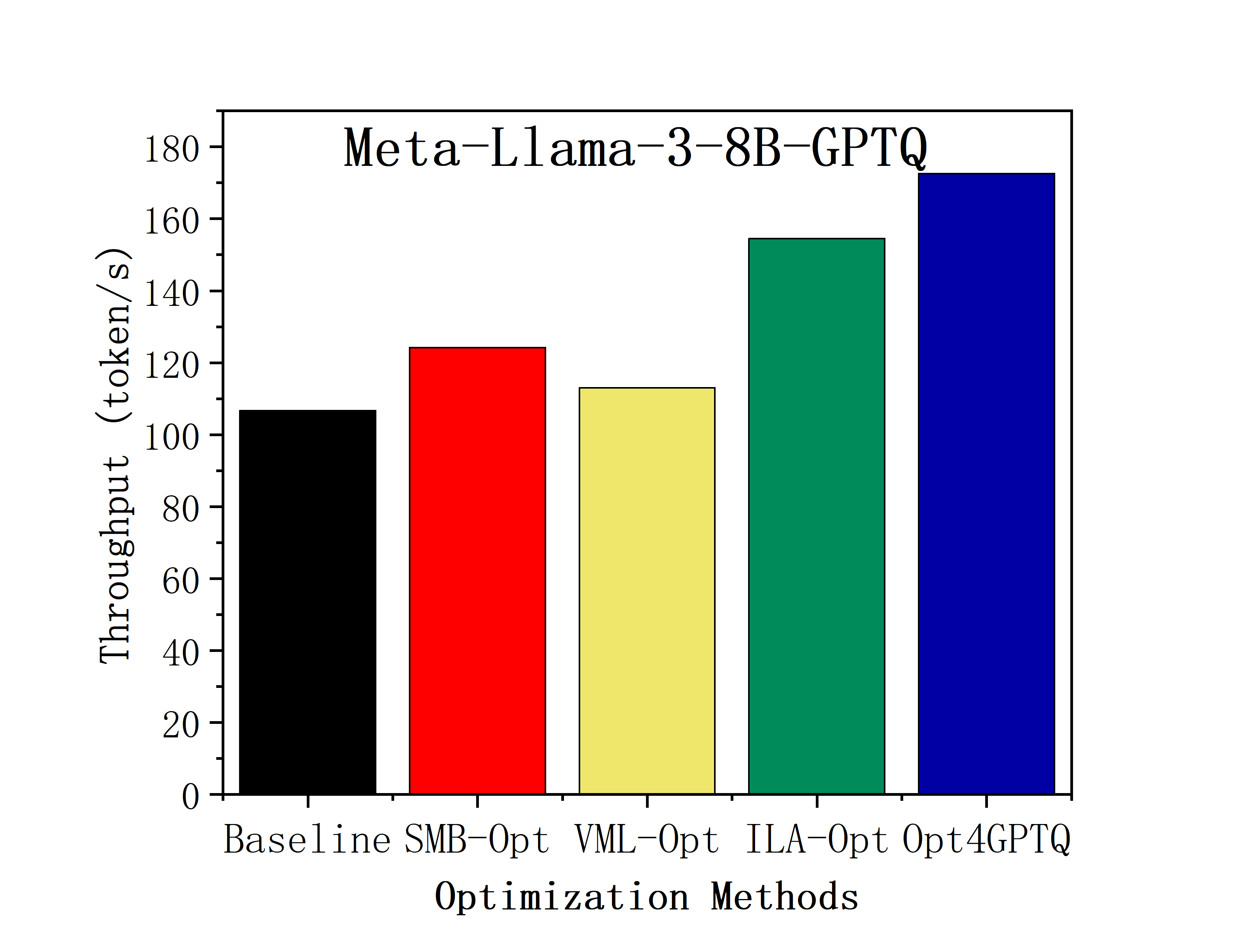}
\caption{Inference throughput comparison of vLLM across different models.}\label{fig2}
\end{figure*}

\begin{table}[h]
\caption{Model throughput improvements with various optimization strategies.}\label{tab3}
\begin{tabular*}{\textwidth}{@{\extracolsep\fill}lcccc}
\toprule
\textbf{Models} & \textbf{SMB-Opt} & \textbf{VML-Opt} & \textbf{ILA-Opt} & \textbf{Opt4GPTQ} \\
\midrule
Meta-Llama-3-8B-GPTQ        & 16.43\% & 5.89\%  & 44.81\% & 61.78\% \\
Llama-2-7B-GPTQ             & 9.50\%  & 4.91\%  & 37.26\% & 54.55\% \\
CodeLlama-7B-GPTQ           & 14.74\% & 5.88\%  & 46.30\% & 67.55\% \\
LLaMa-13B-GPTQ              & 17.98\% & 11.03\% & 57.19\% & 84.42\% \\
Qwen1.5-1.8B-Chat-GPTQ-Int4 & 4.94\%  & 1.36\%  & 16.75\% & 21.93\% \\
Qwen1.5-4B-Chat-GPTQ-Int4   & 6.83\%  & 3.11\%  & 28.74\% & 41.77\% \\
\bottomrule
\end{tabular*}
\end{table}

The throughput evaluation results, illustrated in Figure \ref{fig2} and Table \ref{tab3}, demonstrate that the integrated Opt4GPTQ method significantly boosts vLLM performance, achieving peak gains of 84.42\% and 67.55\% on LLaMa-13B-GPTQ and CodeLlama-7B-GPTQ, respectively. This substantial improvement is driven by the cumulative effects of SMB-Opt and VML-Opt, which reduce memory access overhead and enhance loading efficiency, and ILA-Opt, which provides major instruction-level acceleration via hardware-native vectorized assembly. Notably, larger models exhibit more pronounced gains due to their higher memory and computational intensity, whereas smaller models like Qwen1.5-1.8B-Chat-GPTQ-Int4 show more moderate improvements. These results confirm that synergy between memory optimization and instruction-level acceleration is key to maximizing hardware potential in GPTQ inference.

\subsection{Accuracy Evaluation}
\begin{table}[h]
\caption{Inference accuracy of vLLM on ARC\_C dataset.}\label{tab1}
\begin{tabular*}{\textwidth}{@{\extracolsep\fill}lccccc}
\toprule
\textbf{Models} & \textbf{Baseline} & \textbf{SMB-Opt} & \textbf{VML-Opt} & \textbf{ILA-Opt} & \textbf{Opt4GPTQ} \\
\midrule
Meta-Llama-3-8B-GPTQ         & 75.25\% & 74.92\% & 74.92\% & 74.92\% & 75.25\% \\
Llama-2-7B-GPTQ              & 35.59\% & 36.27\% & 35.25\% & 35.25\% & 35.59\% \\
CodeLlama-7B-GPTQ            & 27.81\% & 28.47\% & 28.47\% & 28.47\% & 29.15\% \\
LLaMa-13B-GPTQ               & 39.32\% & 39.66\% & 39.66\% & 40.00\% & 39.32\% \\
Qwen1.5-1.8B-Chat-GPTQ-Int4  & 48.81\% & 48.81\% & 48.81\% & 48.79\% & 48.81\% \\
Qwen1.5-4B-Chat-GPTQ-Int4    & 56.27\% & 55.59\% & 56.27\% & 56.27\% & 55.59\% \\
\bottomrule
\end{tabular*}
\end{table}

\begin{table}[h]
\caption{Inference accuracy of vLLM on ARC\_E dataset.}\label{tab2}
\begin{tabular*}{\textwidth}{@{\extracolsep\fill}lccccc}
\toprule
\textbf{Models} & \textbf{Baseline} & \textbf{SMB-Opt} & \textbf{VML-Opt} & \textbf{ILA-Opt} & \textbf{Opt4GPTQ} \\
\midrule
Meta-Llama-3-8B-GPTQ         & 87.30\% & 87.48\% & 87.30\% & 87.30\% & 87.30\% \\
Llama-2-7B-GPTQ              & 47.80\% & 47.97\% & 48.59\% & 48.15\% & 47.44\% \\
CodeLlama-7B-GPTQ            & 27.51\% & 27.87\% & 27.87\% & 27.87\% & 27.87\% \\
LLaMa-13B-GPTQ               & 50.79\% & 51.68\% & 51.68\% & 51.50\% & 50.79\% \\
Qwen1.5-1.8B-Chat-GPTQ-Int4  & 69.49\% & 69.14\% & 69.49\% & 69.14\% & 69.14\% \\
Qwen1.5-4B-Chat-GPTQ-Int4    & 70.19\% & 70.19\% & 70.19\% & 70.19\% & 70.19\% \\
\bottomrule
\end{tabular*}
\end{table}

As shown in Table \ref{tab1} and \ref{tab2}, we evaluated the inference accuracy of six models on the ARC\_C and ARC\_E datasets. Across all models and optimization methods, the accuracy variations remain within 1 percentage point, with no consistent upward or downward trends. On ARC\_C, most models exhibit fluctuations within 0.68 percentage points, while on ARC\_E, the largest variation is 0.89 percentage points for LLaMa-13B-GPTQ. Notably, Qwen1.5-4B-Chat-GPTQ-Int4 shows no measurable change. These results indicate that the proposed optimization methods have a negligible impact on inference accuracy, maintaining the original precision with excellent stability.

This numerical stability stems from the fact that our optimizations are implemented at the physical execution layer and are numerically equivalent to the original GPTQ algorithm. Although ILA-Opt utilizes manual inline assembly and SMB-Opt re-orders the accumulation sequence in shared memory, they strictly adhere to the IEEE-754 floating-point standard. The minor observed fluctuations are merely the result of the non-associativity of floating-point addition during parallel reduction, which does not compromise the model's fundamental precision or represent a loss of information.

\section{Conclusion}
This study proposes Opt4GPTQ, a platform-level optimization strategy designed to enhance vLLM inference efficiency on heterogeneous AI acceleration platforms, thereby addressing deployment challenges posed by evolving hardware architectures. Focusing on the GPTQ 4-bit quantized inference within the vLLM serving system, Opt4GPTQ integrates three key techniques: SMB-Opt, VML-Opt, and ILA-Opt. Experimental evaluations demonstrate that Opt4GPTQ significantly boosts throughput across various LLMs while maintaining model accuracy. This work provides practical deployment insights and underscores the necessity of platform-specific engineering for efficient LLM inference. Future work will extend this study by analyzing speedup variations across different batch sizes and evaluating performance with parallel decoding enabled. Moreover, we will validate the generalizability of our method across diverse heterogeneous computing platforms.

\section*{Acknowledgment}
This work was supported in part by the National Supercomputing Center in Jinan, in part by the National Natural Science Foundation of China (NSFC) under Grants 62072287 and W2412090, in part by the HYGON Industry Ecological Cooperation
Organization (HIECO) under Grant 202407027775, and in part by the Shandong Provincial Natural Science Foundation under Grant ZR2024ME230. This paper is an extended version of S4T-GPTQ \cite{b32}, with new contributions in VML-Opt, ILA-Opt, the Opt4GPTQ scheme, and comprehensive inference accuracy evaluations.

\bibliography{wileyNJD-AMA}

\end{document}